\documentclass[preprint,review,a4paper,12pt]{elsarticle}

\usepackage{graphicx}

\usepackage{amssymb}

\biboptions{square}

\journal{Acta Astronautica}

\begin{document}

\begin{frontmatter}


\ead{petri.toivanen@fmi.fi}
\fntext[]{{\it Telephone number}: +358-50-5471521}

\title{Electric sail control mode for amplified transverse thrust}


\author{P. Toivanen, P. Janhunen, and J. Envall}

\address{Finnish Meteorological Institute, FIN-00101, Helsinki, Finland}

\begin{abstract}
The electric solar wind sail produces thrust by centrifugally spanned
high voltage tethers interacting with the solar wind protons. The sail
attitude can be controlled and attitude maneuvers are possible by
tether voltage modulation synchronous with the sail
rotation. Especially, the sail can be inclined with respect to the
solar wind direction to obtain transverse thrust to change the
osculating orbit angular momentum. Such an inclination has to be
maintained by a continual control voltage modulation. Consequently,
the tether voltage available for the thrust is less than the maximum
voltage provided by the power system. Using a spherical pendulum as a
model for a single rotating tether, we derive analytical estimations
for the control efficiency for two separate sail control modes. One is
a continuous control modulation that corresponds to strictly planar
tether tip motion. The other is an on-off modulation with the tether
tip moving along a closed loop on a saddle surface. The novel on-off
mode is introduced here to both amplify the transverse thrust and
reduce the power consumption. During the rotation cycle, the maximum
voltage is applied to the tether only over two thrusting arcs when
most of the transverse thrust is produced. In addition to the
transverse thrust, we obtain the thrusting angle and electric power
consumption for the two control modes. It is concluded that while the
thrusting angle is about half of the sail inclination for the
continuous modulation it approximately equals to the inclination angle
for the on-off modulation. The efficiency of the on-off mode is
emphasized when power consumption is considered, and the on-off mode
can be used to improve the propulsive acceleration through the reduced
power system mass.

\end{abstract}


\end{frontmatter}

\section*{Nomenclature}
\noindent\begin{tabular}{@{}lcl@{}}
A, B &=& thrusting arcs\\
${\bf a}$ &=& acceleration\\
$a$ &=& power series coefficients\\
${\bf e}$ &=& unit vector\\
$F$ &=& force\\
$g$ &=& tether voltage modulation $\in [0,1]$\\
$I$ &=& integral\\
$k$ &=& force parameter\\
$l$ &=& tether length\\
$m$ &=& mass\\
$p$ &=& thrusting arc factor\\
$(r,\theta,\phi)$ &=& spherical polar coordinates\\
${\bf u}$ &=& solar wind velocity\\
$V$ &=& tether voltage\\
(X,Y,Z) &=& Cartesian coordinates\\
$\alpha$ &=& sail angle\\
$\kappa$ &=& force parameter scaled to angular frequency\\
$\Lambda$ &=& sail coning angle\\
$\mu$ &=& free plane tilt angle\\
$\nu$ &=& angular frequency\\
$\rho$ &=& electric sail to centrifugal force ratio\\
$\chi$ &=& $\tan\alpha\tan\Lambda$\\
$\xi$ &=& electric sail thrust factor\\
$\omega$ &=& angular velocity\\
$<>$ &=& angular average\\
$<>_t$ &=& temporal average\\
\end{tabular} \\

\textit{Subscripts} \\
\noindent\begin{tabular}{@{}lcl@{}}
A, B &=& thrusting arcs\\
$a$ &=& angular average\\
$CF$ &=& centrifugal\\
$ES$  &=& electric solar wind sail\\
$(r,\theta,\phi)$ &=& spherical polar coordinates\\
$(x,y,z)$ &=& Cartesian coordinates\\
$\Theta$ &=& thrusting arc index\\
0 &=& reference initial value\\
$\perp$ &=& perpendicular (radial) \\
$\parallel$ &=& parallel (transverse)\\
$*$ &=& orbital coordinates\\
\end{tabular} \\

\textit{Superscripts} \\
\noindent\begin{tabular}{@{}lcl@{}}
$\Theta$ &=& thrusting arc index\\
\end{tabular} \\


\section{Introduction}
\label{sec:intro}

\begin{figure}[htb]
	\centering\includegraphics[width=2.5in]{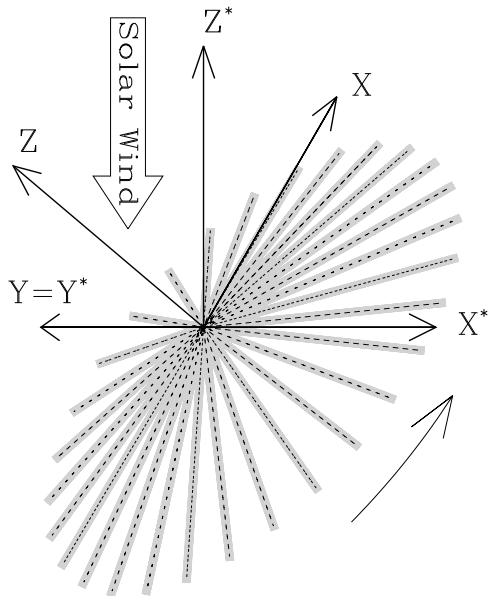}
	\caption{Slowly rotating electric solar wind sail with an
          electrostatic effective sail area (gray shading) around the
          thin tethers (dashed lines).}
	\label{fig:overview}
\end{figure}

The thrust of the electric solar wind sail is produced through an
interaction of the solar wind protons and electrostatic electric field
of long electrically charged tethers \cite{1}. The tethers are spanned
centrifugally to form a sail rig slowly rotating together with the
spacecraft main body (Fig. \ref{fig:overview}). The positive
tether voltage of a few tens of kilovolts is actively maintained by an
electron gun powered by solar panels. The spatial scale size of the
electric field structure around the tethers is several hundreds of
meters forming an effective sail area against the solar wind dynamic
pressure. The obtained thrust is several hundreds of nN/m over the
tether length \cite{2}. The tethers are light-weight and made of
micrometer thin (a few tens of $\mu$m) aluminum wires ultrasonically
bonded together \cite{3} for redundancy against the
micro-meteoroid flux.

To maintain the tethers rotating in unison, there are two principal
electric solar wind sail designs. One assumes mechanically coupled
tethers by flexible auxiliary tethers connecting the main tether tips
\cite{4}. At each tether tip, there is a remote unit that includes
auxiliary tether reels for the sail deployment while the main tethers
are reeled out from the central body of the spacecraft. As a baseline,
miniature cold gas thrusters are also included to start the sail
rotation by producing the required angular momentum \cite{5}. The
other design assumes that the tethers are mechanically uncoupled and
the rotation rate is controlled by freely guided photonic blades at
each tether tip \cite{6}.

The electric solar wind sail attitude maintenance and maneuvers can be
introduced by modulating the voltage of each tether individually and
synchronously with the spacecraft rotation. The sail nominal rotation
plane can be turned to incline the sail thrust vector relative to the
Sun-spacecraft direction. In the normal flying mode, the sail is
inclined with respect to the Sun-spacecraft direction either to gather
or diminish the osculating orbit angular momentum. In this mode,
tether voltages have to be modulated continually. As the average
tether voltage is less than the maximum provided by the spacecraft
power system, the control mode affects the sail efficiency.

\begin{figure}[htb]
	\centering\includegraphics[width=2.5in]{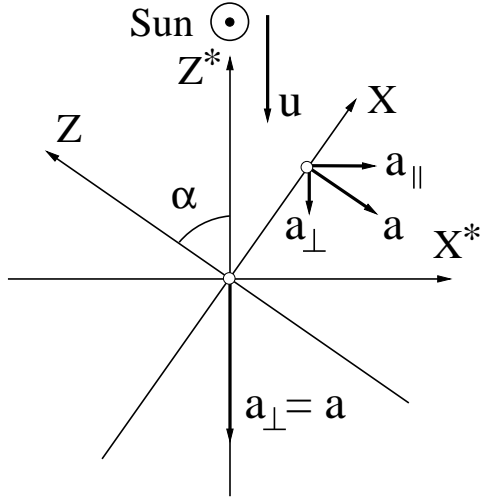}
	\caption{Geometry and rotation phase dependence of the thrust
          orientation in the sail coordinates (X, Z) and orbital
          coordinates (X$^*$, Z$^*$).}
	\label{fig:coords_a}
\end{figure}

The electric solar wind sail thrusting geometry is such that the
thrust generated by a single tether is along the solar wind component
perpendicular to the tether (Fig. \ref{fig:coords_a}). For simplicity,
the coordinate systems in this study are as follows. The spacecraft
orbital coordinates are such that X$^*$ is along the spacecraft orbital
velocity, Z$^*$ points to the Sun, and Y$^*$ is normal to the orbital
plane. The sail coordinates (X,Y,Z) are then rotated by the sail angle
($\alpha$) around the Y$^*$ axis as shown in
Fig. \ref{fig:coords_a}. The thrust magnitude and direction depend
then on the tether rotation phase as depicted in
Fig. \ref{fig:coords_a} in terms of the tether acceleration (${\bf
  a}$). When the tether (white circle) is normal to the orbital plane
(X-Z plane) the thrust has only a radial component (${\bf
  a}_{\perp}$). Maximum transverse thrust (${\bf a}_{\parallel}$) is
obtained when the tether is in the orbital plane. The total thrust
vector is then obtained as a rotation phase average together with the
control voltage modulation combining the effects of the thrust
geometry and the voltage modulation to the overall sail thrust.

\begin{figure}[htb]
	\centering\includegraphics[width=2.5in]{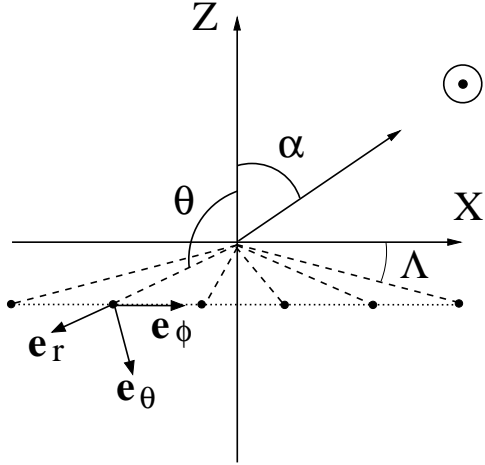}
	\caption{Number of tethers (dashed lines) in the sail
          coordinates with their attitude defined by the sail
          ($\alpha$) coning ($\Lambda$) angles, and the spherical
          coordinate unit vectors (${\bf e}_r, {\bf e}_{\theta}, {\bf
            e}_{\phi}$).}
	\label{fig:coords_b}
\end{figure}

Recently, the electric solar wind sail tether was modeled as a
spherical pendulum \cite{13}. As the tether is much longer (up to a
few tens of kilometers) than any spacecraft spatial size, the central
plate effect can be neglected, and the tether is rather a spherical
than a rotating pendulum (pendulum pivot attached to a rotating
plate). It was shown that there is an analytical form for the voltage
modulation that maintains the sail attitude with respect to any
practical Sun direction (Fig. \ref{fig:coords_b}). As a result, the
tether rotates in a cone with its tip rotating in a plane. The coning
angle is defined by the electric solar wind sail and centrifugal
forces.

The sail guidance, including the thrust vectoring, spin plane
maneuvers, attitude maintenance, and navigation in variable solar wind
is a key component in mission analysis of the electric solar wind sail
applications. Given the baseline of 1 N thrust of a full scale sail
\cite{massmodel}, several types of missions have been suggested and
analyzed \cite{applications}, \cite{5}. These include outer solar
system exploration\cite{outersolarsystem}, missions in non-Keplerian
orbits\cite{nonkeplerian}, asteroid
missions\cite{neo},\cite{asteroids}, and protection from hazardous
asteroids\cite{asteroids},\cite{sini}. Many of these missions require
accurate navigation to the target in variable solar wind
conditions\cite{navigation}. Based on the results of this paper, the
thrust vectoring and sail control mode efficiency in terms of the
transverse thrust and electric power consumption are available for the
future sail navigation and mission analysis.

In this paper, the motion of an on-off controlled tether is analyzed in
Section \ref{sec:onoffanalysis}. The efficiencies of the smooth and
on-off and control modes are obtained in Section
\ref{sec:modeanalysis}. The control modes are then compared in Section
\ref{sec:results} in terms of the transverse thrust, electric power
consumption, and thrusting angle.

\section{Analysis of tether motion for on-off mode}
\label{sec:onoffanalysis}

\subsection{Equations of motion}
\label{subsec:eqmotion}

The tether dynamics with a fixed tether length can be described as
a spherical pendulum with equations of motion
\begin{eqnarray}
a_{\theta} &=& \ddot{\Lambda} + \cos\Lambda\sin\Lambda\,\dot{{\phi}}^2 = -gk\left(\sin\alpha\sin\Lambda\cos\phi+\cos\alpha\cos\Lambda\right)\label{eq:lambdamotion}\\
a_{\phi} &=& \cos\Lambda\ddot{\phi} - 2\sin\Lambda\dot{\Lambda}\dot{\phi} = - gk\sin\alpha\sin\phi.
\label{eq:phimotion}
\end{eqnarray}
The acceleration ($a_{\theta},a_{\phi}$) is given in spherical
coordinates $(r,\theta,\phi)$ corresponding to the sail coordinates as
defined in Figs. \ref{fig:coords_a} and \ref{fig:coords_b}. The
equations of motion are given in terms of the coning angle $\Lambda =
\theta - \pi/2$. Assuming that the tether has a constant linear mass
density, the tether motion is parametrized by
\begin{equation}
k = \frac{3\xi u}{2 m},
\label{eq:k}
\end{equation}
where $\xi$ is constant arising from the electric sail thrust law
\cite{2}, $u$ is the solar wind velocity, and $m$ is the tether
mass. Note that $k$ is negative as the nominal solar wind ($u$) is
flowing antiparallel with Z$^{*}$ axis. The tether voltage modulation
is given by $g \in [0,1]$.

\begin{figure}[htb]
	\centering\includegraphics[width=2.5in]{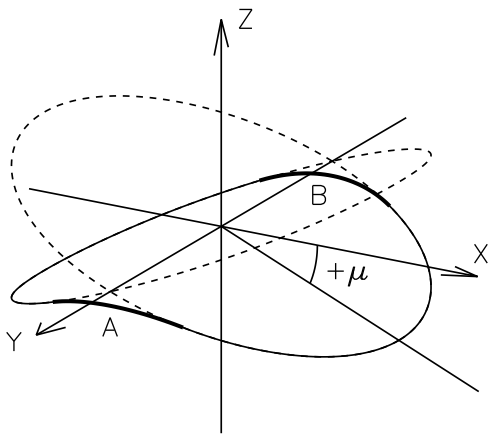}
	\caption{Tether tip trajectory (solid curve) in sail coordinates
          for on-off control mode. Two thrusting arcs A and B
          ($g=1$, thick solid curve) connects the free planar motion
          ($g=0$) on two planes (dashed lines) inclined by $\pm \mu$.}
	\label{fig:taco}
\end{figure}

Our aim is to seek a piecewise solution for the equation of motion as
sketched in Fig. \ref{fig:taco}. The solution consists of two sections
of free planar motion ($g=0$, solid curve) connected by two thrusting
arcs ($g=1$, thick solid curve). The solution is given as
$\Lambda(\phi)$ and $\dot\phi(\phi)$ separately for the free motion
and the forced thrusting motion. The forced motion is solved as a
power expansion in $\phi$, and its free parameters are fixed by the
continuity conditions at the limits of thrusting arcs.

\subsection{Free tether motion}
\label{subsec:freemotion}

When the tether voltage is off ($g = 0$), the tether rotates in a plane with angular speed $\omega$ with
\begin{equation}
\omega^2 = \cos^2\Lambda\dot\phi^2 + \dot\Lambda^2.
\label{eq:omega2}
\end{equation}
The angular speed is a constant of free motion. The equation of the
 plane can be given as
\begin{equation}
\tan\Lambda + \tan\mu\sin\phi = 0,
\label{eq:freeplane}
\end{equation}
where we have assumed that the plane is tilted around $Y$ axis by
angle $\mu$ as shown in Fig. \ref{fig:taco}. Furthermore,
Eq. (\ref{eq:phimotion}) can be written as
\begin{equation}
\frac{\rm d}{\rm dt}(\cos^2\Lambda\dot\phi) = 0.
\label{eq:omegaz}
\end{equation}
implying that the Z component of the angular velocity is a constant of
motion. The free motion is fixed ($\omega \equiv \omega_0$) at
$(X,Y,Z) = (0,l\cos\mu,l\sin\mu)$, where $\phi = \pi/2$, $\Lambda =
\mid \mu \mid$, and $\dot\Lambda = 0$. Using Eq. (\ref{eq:omega2}) at this
point, the constant associated with Eq. (\ref{eq:omegaz}) can be fixed,
and $\cos^2\Lambda\dot\phi = \omega_0\cos\mu$. Finally, solving
$\cos^2\Lambda$ using Eq. (\ref{eq:freeplane}), the square of the
angular frequency can be written as
\begin{equation}
\dot\phi^2 = \frac{\omega_0^2(1+\tan^2\mu\sin^2\phi)^2}{1+\tan^2\mu}.
\label{eq:freedotphi}
\end{equation}

\subsection{Forced tether motion}
\label{subsec:forzedmotion}

When the voltage is turned on the tether leaves the initial plane of
free motion and adopts another plane when the voltage is turned off
again. Especially, one can first consider two planar tether tip orbits
as shown with dashed curves in Fig. \ref{fig:taco}. When the tether
voltage is turned on for the sector (thrusting arc A) near the
positive X axis, the tether tip transits from one plane to the other
($\mu \rightarrow -\mu$). Turning the voltage on around negative X
axis (arc B), the tether returns to the initial plane. In \ref{app:a},
we show that such trajectories exist. For given arc length A
($\phi_A$), electric solar wind sail force parameter ($k$), and tether
initial angular speed ($\omega_0$), the free plane tilt angle
\begin{equation}
\tan\mu = \kappa_0\cos\alpha\phi_A(1-2\kappa_0\sin\alpha\phi_A^2 + \mathcal{O}{(\phi^4)}),
\end{equation}
and arc length B
\begin{equation}
\phi_B = \phi_A(1-4\kappa_0\sin\alpha \phi_A^2 + \mathcal{O}{(\phi^4)})
\label{eq:phiB_s}
\end{equation}
can be solved to define the tether motion under the on-off
modulation. Note that the arcs are nonsymmetric due to the sail
inclination, and $\kappa_0 = k/\omega_0^2$.

\begin{figure}[htb]
	\centering\includegraphics[width=2.8in]{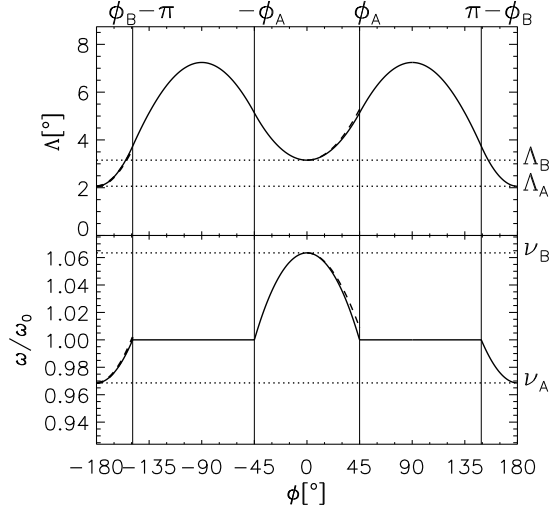}
	\caption{Analytical (solid) and numerical (dashed) solutions
          for the tether coning angle $\Lambda$ (upper panel) and
          angular velocity (lower panel) as functions of azimuthal
          angle $\phi$ with $\alpha$ = 45$^{\circ}$.}
	\label{fig:lambdaetomega}
\end{figure}

Fig. \ref{fig:lambdaetomega} shows the tether local coning angle and
angular velocity as functions of azimuthal angle $\phi$. The largest
descent of the tether from X-Y plane is at Y axis corresponding to
free plane inclination $\mu$. The closest approaches are at X axis as
determined by $\Lambda_A$ and $\Lambda_B$ in
Eq. (\ref{eq:LambdaT}). The asymmetry in thrusting arcs A and B is due
to the sail angle being 45$^{\circ}$. The asymmetry is also reflected
in the minimum and maximum angular velocities as given by $\nu_A$ and
$\nu_B$ in Eq. (\ref{eq:nua}). The analytical solutions are well
justified against the numerical solution for the thrusting arcs A and
B as shown by the dashed curves. Note that the deviation of the
analytical solution from the numerical one is mostly due to the fact
that the thrusting arc half length $\phi_A$ being 45$^{\circ}$ is
pushing the limits of the approximation based on a power series
expansion in $\phi$.

\section{Analysis of control mode efficiencies}
\label{sec:modeanalysis}

The Cartesian components of the acceleration in terms of the spherical
components read as
\begin{eqnarray}
a_x &=& -a_{\theta}\sin\Lambda\cos\phi - a_{\phi}\sin\phi \label{eq:cartesianax}\\
a_y &=&- a_{\theta}\sin\Lambda\sin\phi + a_{\phi}\cos\phi \label{eq:cartesianay}\\
a_z &=& - a_{\theta}\cos\Lambda
\label{eq:cartesianaz}
\end{eqnarray}
that are given in terms of the coning angle $\Lambda$ as in
Eq. (\ref{eq:lambdamotion}) and Eq. (\ref{eq:phimotion}). The
acceleration is calculated in the sail coordinates and must be rotated
by the sail angle to the orbital coordinates for the radial and
transverse thrust components.

To compare the control mode efficiencies, it is assumed that the
tether is initially rotating freely with $\omega = \omega_0$ before
any control mode is applied. Then we define a parameter $\rho$ as a
ratio of the electric solar wind sail ($F_{ES} = \xi \mid u \mid l$)
and the centrifugal force ($F_{CF} = 1/2 m \omega_0^2 l$) integrated
over the tether length $l$,
\begin{equation}
\rho = \frac{F_{ES}}{F_{CF}} = \frac{2 \xi \mid u\mid}{m \omega_0^2} = \frac{4}{3}\frac{\mid k\mid}{\omega_0^2}.
\label{eq:rho}
\end{equation}
Use of $\rho$ is motivated by the fact that the coning angle
($\lambda$) is not a natural parameter for the on-off mode tether
attitude ($\dot\lambda \neq 0$) as it is for the smooth mode
($\dot\lambda = 0$).

The electric power for the electric solar wind sail tether depends on
the tether voltage $V$ and is proportional to $V^{3/2}$ \cite{14}. Thus
the power estimate is then proportional to the temporal average of
$g^{3/2}$ over the rotation period ($<g^{3/2}>_t$). For simplicity,
however, we consider the angular average $<g>^{3/2}$ as the power
estimate. Numerical calculations show that this simplification
underestimates the power consumption of the smooth mode by a few
percent ($< 4\%$) depending on the sail plane inclination and the
ratio of sail and centrifugal forces.

\subsection{Smooth mode efficiency}
\label{sub:smootheffi}

Using the equation of motion, it can be shown that modulation
\begin{equation}
g=\left(\frac{1-\chi}{1+\chi\cos\phi}\right)^3
\label{eq:modulation}
\end{equation}
keeps the sail coning angle constant \cite{13}, where $\chi =
\tan\alpha\tan\Lambda$. Note also that the control signal is
normalized to its maximum value (max$(g) = 1$) instead of $<g> = 1$
\cite{13}.

\begin{figure}[htb]
	\centering\includegraphics[width=2.8in]{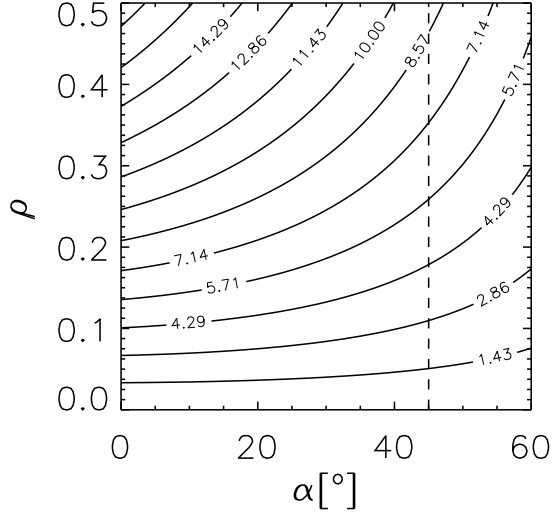}
	\caption{Coning angle for smooth modulation}
	\label{fig:coning}
\end{figure}

When the smooth modulation is turned on the tether proceeds to rotate
with some coning angle $\Lambda$. During a slow transition,
Eq. (\ref{eq:omegaz}) holds, $\omega_Z$ is an adiabatic invariant, and
\begin{equation}
\nu_a\cos^2\Lambda = \omega_0,
\label{eq:omegazo}
\end{equation}
where $\nu_a$ is the angular average of the angular frequency
($<\dot\phi>$). Furthermore, substituting Eq. (\ref{eq:modulation}) in
Eq. (\ref{eq:lambdamotion}), averaging over the rotation phase
($\ddot\Lambda = 0$), and using Eqs. (\ref{eq:rho}) and
(\ref{eq:omegazo}), we write
\begin{equation}
\rho = \frac{4\sin\Lambda}{3\cos\alpha\cos^4\Lambda}\frac{(1-\chi^2)^{\frac{3}{2}}}{(1-\chi)^3}
\end{equation}
in terms of the coning angle $\Lambda$ ($\chi =
\tan\alpha\tan\Lambda$). The integral associated with the averaging is
given by Eq. (\ref{eq:integrala}) of \ref{app:b}. The coning angle
$\Lambda$ can then be solved numerically as a function of $\rho$ as shown
in Fig.  \ref{fig:coning}.

To calculate the thrust, the spherical components of the acceleration
of Eqs. (\ref{eq:lambdamotion}) and (\ref{eq:phimotion}) are replaced
by Eqs. (\ref{eq:cartesianax})--(\ref{eq:cartesianaz}). Non-zero
components of the rotation phase averaged acceleration components can
be written as
\begin{eqnarray}
<a_x> &=& k(\sin\alpha<g> + \cos\alpha\cos\Lambda\sin\Lambda <g\cos\phi> \nonumber\\
&-& \sin\alpha\cos^2\Lambda <g\cos^2\phi>)\\
<a_z> &=& k\cos\alpha\cos^2\Lambda\,(<g> + \chi<g\cos\phi>).
\end{eqnarray}
after using trigonometric identities. 

The phase averaged quantities are of the form of $<g\cos^n\phi> =
(1-\chi)^3I_n$ where
\begin{equation}
I_n = \frac{1}{2\pi}\int_0^{2\pi}\frac{\cos^n\phi d\phi}{(1+\chi\cos\phi)^3},\,\,\, n = 0, 1, 2.
\end{equation}
These integrals can be executed by a computer algebra system such as
Maxima\cite{maxima} and are given in \ref{app:b}. The average modulation
$<g>$ needed for the power consumption estimate reads as
\begin{equation}
<g> = (1-\chi)^3I_0 = \frac{(1-\chi)^3(2+\chi^2)}{2(1-\chi^2)^{\frac{5}{2}}}.
\label{eq:aveg_a}
\end{equation}
After some manipulation, the phase averaged acceleration components
read as
\begin{eqnarray}
 <a_x> &=& \frac{k}{2}\sin\alpha\cos 2\Lambda\,\frac{(1-\chi)^3}{(1-\chi^2)^{\frac{3}{2}}}\\
 <a_z> &=& k\cos\alpha\cos^2\Lambda\,\frac{(1-\chi)^3}{(1-\chi^2)^{\frac{3}{2}}}.
\end{eqnarray}
The Y component of the acceleration includes terms proportional either
to $<g\sin\phi>$ or $<g\sin\phi\cos\phi>$ that equal to zero. Finally,
these components are rotated to the orbital coordinates giving the
radial and transverse thrust as
\begin{eqnarray}
a_{\perp} &=& \frac{k}{2}\frac{(1-\chi)^3}{(1-\chi^2)^{\frac{3}{2}}}(2\cos^2\Lambda - \sin^2\alpha)\label{eq:radial_a}\\
a_{\parallel} &=& - \frac{k}{4}\frac{(1-\chi)^3}{(1-\chi^2)^{\frac{3}{2}}}\sin 2\alpha   
\label{eq:trans_a}
\end{eqnarray}
The thrust angle can be expressed by
\begin{equation}
\tan\psi = \frac{a_{\parallel}}{a_{\perp}} = -\frac{\sin2\alpha}{2(\cos^2\Lambda-\sin^2\alpha)}.
\end{equation}

\subsection{On-off mode efficiency}
\label{sub:onoffeffi}

\begin{figure}[htb]
	\centering\includegraphics[width=2.8in]{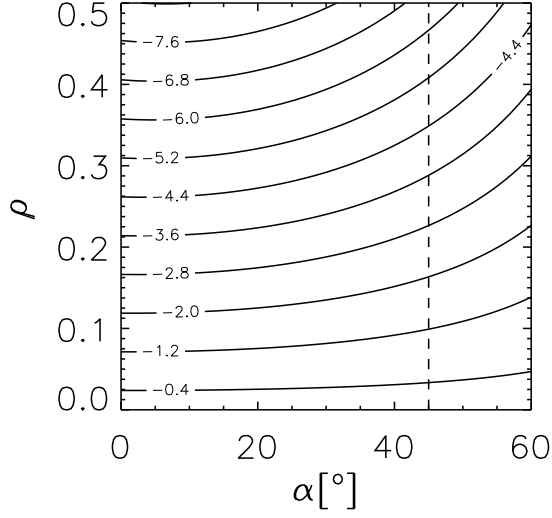}
	\caption{Isocontours of free plane tilt angle in degrees for
          $\phi_A = 22.5^{\circ}.$}
	\label{fig:mu}
\end{figure}

For the on-off mode, the tether attitude can be described by the tilt
angle of the free plane ($\mu$) as a function of $\rho$ and $\alpha$
(Fig. \ref{fig:mu}). This corresponds to the the maximum instantaneous
tether coning angle and is about half of the smooth mode coning angle
shown in Fig. \ref{fig:coning}.

The efficiency analysis of the on-off mode can be carried out by using
the results of Section \ref{subsec:forzedmotion}. The average
modulation for the power consumption can be written as
\begin{equation}
<g> = \frac{1}{\pi}(2\phi_A + \delta\phi) \approx
\frac{2}{\pi}\phi_A(1 - 2\kappa_0\sin\alpha\phi_A^2 +
\mathcal{O}{(\phi_A^4)}).
\label{eq:aveg_b}
\end{equation}
Similarly to the smooth mode efficiency analysis, using
Eqs. (\ref{eq:lambdamotion}), (\ref{eq:phimotion}), (\ref{eq:cartesianax}),
(\ref{eq:cartesianay}), (\ref{eq:cartesianaz}), and linearizing in
$\Lambda$, the averaged components of acceleration can be written as
\begin{eqnarray}
<a_x> &=&k(\cos\alpha<g\Lambda\cos\phi> +\sin\alpha<g\sin^2\alpha>)\\
<a_z> &=&k(\sin\alpha<g\Lambda\cos\phi> +\cos\alpha).
\end{eqnarray}
Expanding these components in $\phi$, integrating over the arcs A and
B, and rotating the acceleration components to the orbital
coordinates, the average radial and transverse components of the
acceleration can be written as
\begin{eqnarray}
a_{\perp} &=& \frac{2k}{\pi}\phi_A(\cos^2\alpha \nonumber\\&+& \textsc{\small 1/}_{\textsc{\small 3}}\,(\sin^2\alpha - 10\kappa_0\sin\alpha\cos^2\alpha) \phi_A^2 + \mathcal{O}{(\phi_A^4)})\label{radial_b}\\
a_{\parallel} &=& -\frac{k}{\pi}\phi_A(\sin 2\alpha \nonumber\\&-& \textsc{\small 1/}_{\textsc{\small 3}}\,(\sin2\alpha - 4\kappa_0\cos^3\alpha(1 - 4\tan^2\alpha)) \phi_A^2 + \mathcal{O}{(\phi_A^4)})
\label{eq:trans_b}
\end{eqnarray}
The thrusting angle of the on-off modulation can be expressed then as
\begin{eqnarray}
\tan\psi &=& -\tan\alpha \nonumber\\&+&\textsc{\small 1/}_{\textsc{\small 3}}\,(1+\tan^2\alpha)(\tan\alpha-2\kappa_0\cos\alpha)\phi_A^2 + \mathcal{O}{(\phi_A^4)},
\end{eqnarray}
where the leading term is $-\tan\alpha$ as expected. The minus sign is
due to the definition of $a_{\perp}$ being negative for all sail
angles.

\section{Results: control mode comparison}
\label{sec:results}

\subsection{Power consumption}
\label{sub:powerconsumption}

\begin{figure}[htb]
	\centering\includegraphics[width=2.8in]{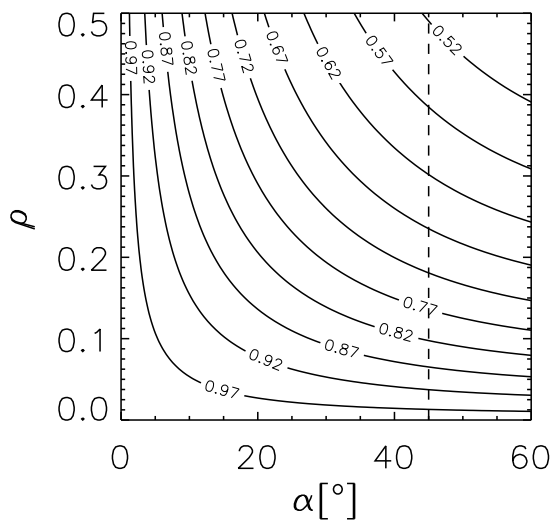}
	\caption{Power consumption of smooth modulation.}
	\label{fig:power_a}
\end{figure}

\begin{figure}[htb]
	\centering\includegraphics[width=2.8in]{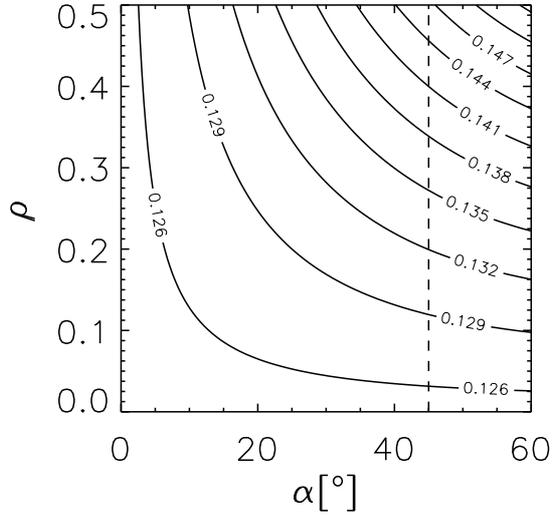}
	\caption{Power consumption of on-off modulation with $\phi_A =
          22.5^{\circ}$.}
	\label{fig:power_b}
\end{figure}

The estimates for the power consumption of the control modes were
given by Eq. (\ref{eq:aveg_a}) and Eq. (\ref{eq:aveg_b}). Fig.
\ref{fig:power_a} shows power consumption for the smooth
modulation. It can be seen that the control mode efficiency falls
considerably as a function of $\rho$ and sail angle. For the on-off
modulation, the power consumption is about 0.12-0.15 times of the
maximum of the smooth modulation as shown in Fig.
\ref{fig:power_b}. Note that this depends on the thrusting arc
length. Weak dependence on $\rho$ and sail angle arises from the fact
that for non-zero sail angle the thrusting arcs are different in
length according to Eq. (\ref{eq:deltaphi}).

\subsection{Transverse thrust}
\label{sub:transverse}

\begin{figure}[htb]
	\centering\includegraphics[width=2.8in]{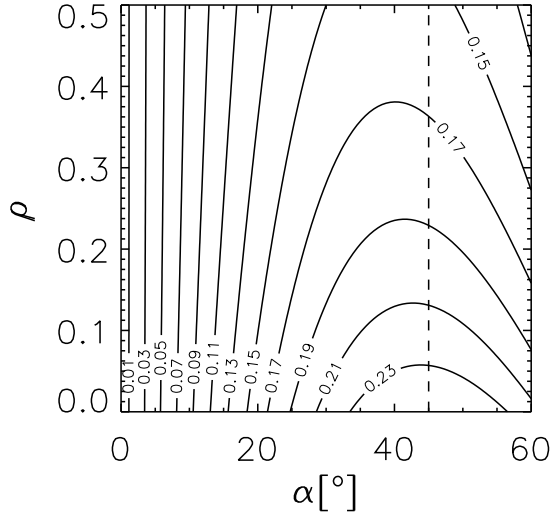}
	\caption{Transverse thrust for smooth modulation scaled to the maximum total thrust.}
	\label{fig:trans_a}
\end{figure}

\begin{figure}[htb]
	\centering\includegraphics[width=2.8in]{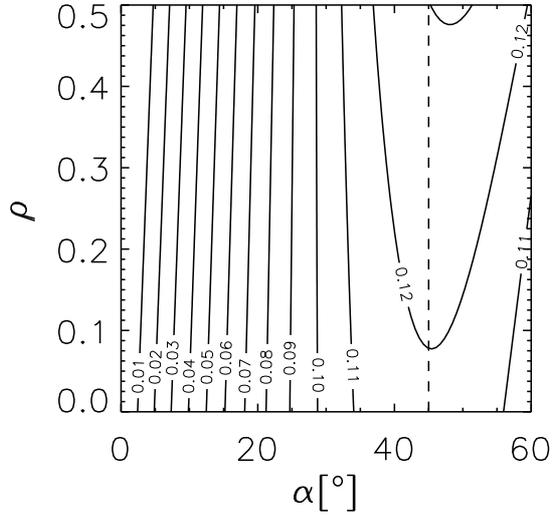}
	\caption{Transverse thrust for on-off modulation}
	\label{fig:trans_b}
\end{figure}

Another important parameter to characterize the control modes is the
amount of transverse thrust. Figs. \ref{fig:trans_a} and
\ref{fig:trans_b} show the transverse thrust for smooth and on-off
modulations as obtained from Eq. (\ref{eq:trans_a}) and
Eq. (\ref{eq:trans_b}), respectively. The transverse thrust is scaled
to the maximum total thrust (1) obtained for the smooth modulation
with fast rotating sail and zero sail angle ($\rho$, $\alpha$) =
(0,0). It can be seen that the transverse thrust is about 25\% of the
total thrust for the smooth modulation. In addition, the transverse
thrust falls significantly both in $\alpha$ and $\rho$ from its
maximum. For example, the transverse thrust is about 17\% when $\rho
\approx 3.5$, corresponding to a coning angle of 7$^{\circ}$. Note
that the same behavior can be obtained by the sail angle of about
20$^{\circ}$ for a fast spinning sail (zero coning angle). These
results indicates that the stronger the tether material is, the
smaller fraction of available power is required for the sail attitude
control, and the more efficient the sail is.

For the on-off modulation, the transverse thrust is about 10\% for
properly inclined sail. Note that it is expected that the transverse
thrust of the on-off modulation is smaller than that given by smooth
modulation since a fraction of the transverse thrust comes outside the
thrusting arc. Another way to compare the transverse thrust is to
scale the control mode transverse thrust by the power system mass that
is proportional to the power consumption. The scaling factor is
about 0.15 as the maximum of the power consumption of the on-off mode
shown in Fig. \ref{fig:trans_b}. Thus the maximum scaled transverse
thrust is about 0.8 (= 0.12/0.15) to be compared with 0.23 of the smooth
mode (Fig. \ref{fig:trans_b}).

\subsection{Thrust angle}

\begin{figure}[htb]
	\centering\includegraphics[width=2.8in]{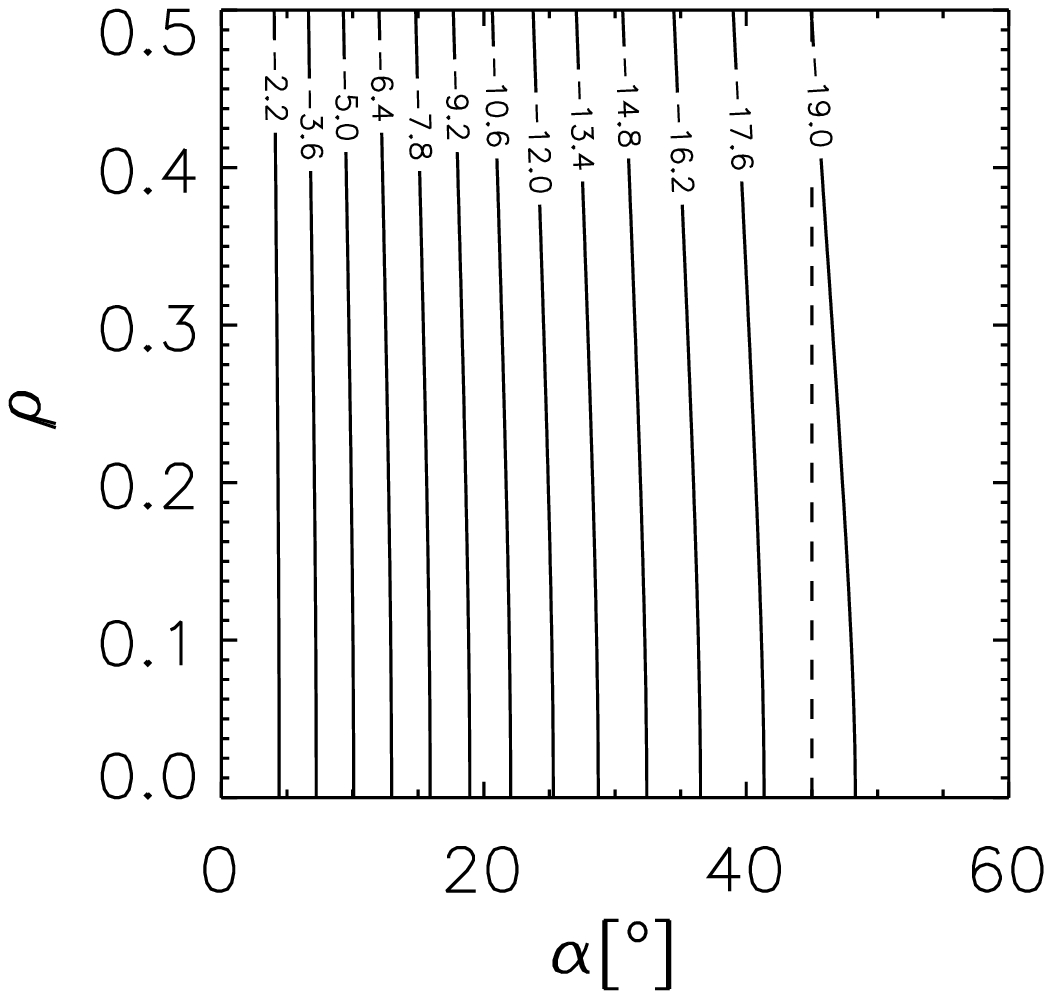}
	\caption{Thrust angle for smooth modulation}
	\label{fig:thrusta_a}
\end{figure}

\begin{figure}[htb]
	\centering\includegraphics[width=2.8in]{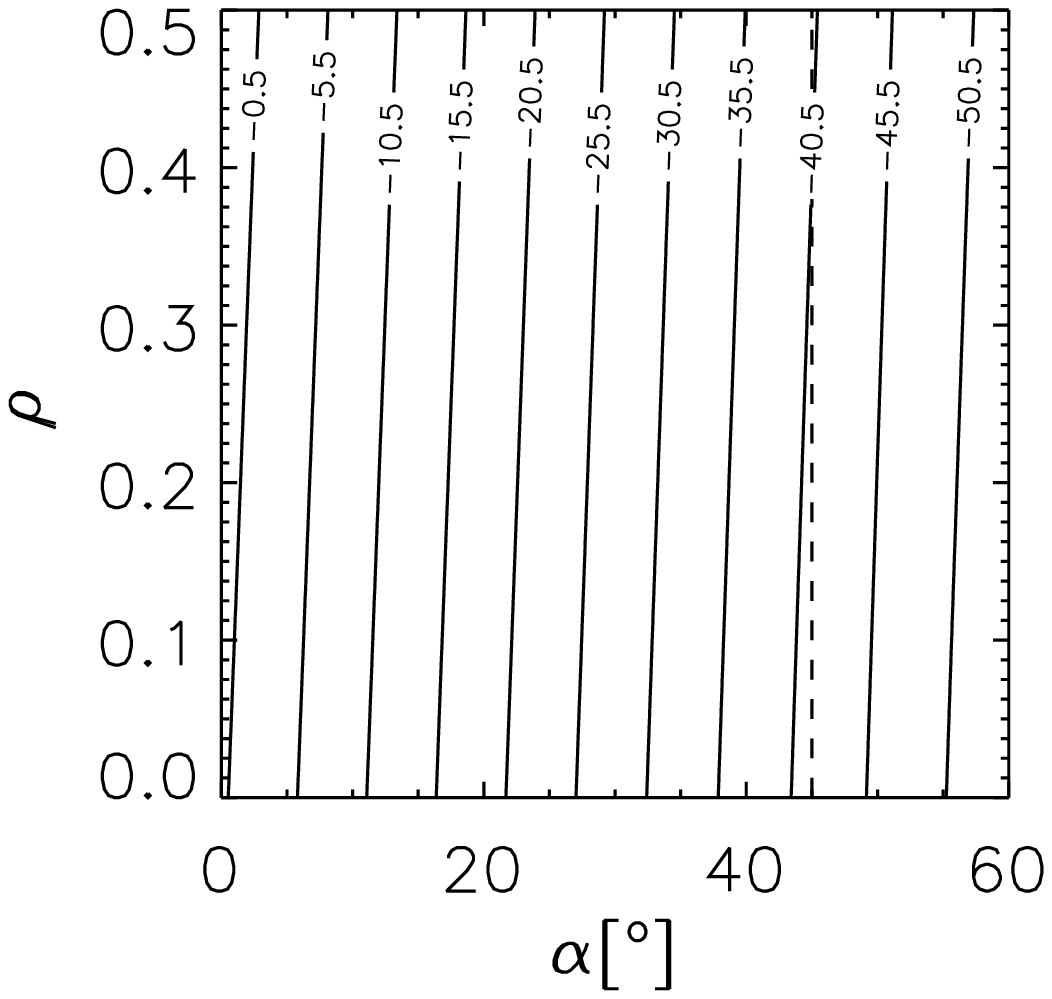}
	\caption{Thrust angle for on-off modulation}
	\label{fig:thrusta_b}
\end{figure}

To compare the radial and transverse thrust for the control modes,
Figs. \ref{fig:thrusta_a} and \ref{fig:thrusta_b} show the thrust
angle. In the case of the smooth modulation, the thrust angle is about
half of the sail angle and reaches its maximum (about 20$^{\circ}$)
when the sail angle of 45$^{\circ}$. There is no use of turning the
sail more than 45$^{\circ}$ in terms of the thrust angle when using
the smooth modulation. For the on-off modulation, the thrust angle
roughly equals to the sail angle.

\section{Discussion}
\label{sec:discussion}

In this work, we assumed that the tether dynamics can be described by
a spherical rigid rod pendulum. The single pendulum that is
mechanically uncoupled to the neighboring tethers describes the
electric solar wind sail design assuming freely guided photonic blades
for the additional spin rate control. The ultimate difference in
tether dynamics of this design is that the tether rotation rate varies
in rotation phase, especially when the smooth modulation is used. This
implies a voltage modulation amplitude larger than that required for
the mechanically coupled design. Thus the smooth mode efficiency is
underestimated in this study for the auxiliary tether design with the
tethers rotating in unison. In the case of the on-off modulation, the
tether rotation rate variation is negligible, and it can be expected
that the estimates given here are valid for the auxiliary tether
design when using the on-off modulation. For better analytical
estimates for the smooth modulation, the electric solar wind sail can
be assumed as a rigid body with a number of tethers connected by the
auxiliary tethers as a future study.

In practice, realistic tether voltage finite rise and decay times
have to be taken into account. This affects both the tether control
and the efficiency estimates. We tested a simple control routine to
see that the tether tip can be maintained on a closed loop resembling
closely the motion of the on-off modulated tether described here. The
scheme was such that the sail coordinates were rotated back (and
forth) corresponding to the flip of the free rotation plane while the
tether was moving along arcs A (and B). In these rotated coordinates,
we controlled the temporal change of the tether local coning angle to
keep the tether rotating in (X,Y) plane of the rotated coordinates
system. The scheme resulted in a smoothed on-off tether modulation
signal. As the scheme is far from being technically feasible for the
tether control as such the result were not discussed in detailed in
this paper. We anticipate however that the efficiency estimates do not
change dramatically due to the on-off modulation with realistic rise
and decay times.

In reality, the tethers are flexible and the sail cannot be rotated
infinitely fast the tethers to behave as rigid rods under the
centrifugal force. Thus it can be expected that the on-off modulation
may cause some undesired slow oscillations of the tether, especially
if the rotation frequency meets the tether oscillation eigenmode
frequency. This issue has to be addressed by a fully dynamical
simulation with a flexible tether. At this stage, it can be argued
that these frequency ranges can be designed not be in resonance with
each other. Furthermore, the realistic tether voltage rise and decay
times smooths out transients that may be causes for harmful tether
oscillations.

The real flexible tethers also implies that the local coning angle
decreases towards the tip of the rotating tether due to the enhancing
centrifugal force along the tether. The effective coning angle of a
flexible tether is then less than that obtained from the ratio of the
electric sail and the centrifugal forces for a rigid rod, and the
results here underestimate the efficiency of the smooth control
mode. However, solving the actual shape and effective coning angle of
the flexible tether as a future study, the results shown here are
applicable by replacing the coning angle used with the effective
coning angle.

\section{Conclusions}
\label{sec:summary}

Two electric solar wind sail attitude control schemes were analyzed in
this paper. One assumes smooth tether voltage modulation consistent
with strictly planar tether tip motion. The other is based on an
on-off voltage modulation leading to a non-planar but closed loop
tether tip motion. The tether motion of the former mode is based on
earlier results while the solution for the tether tip motion of the
latter mode was derived here in details. Having the dynamics of the
both modes solved, the efficiency of the control modes was addressed
in terms of the sail transverse thrust, thrust angle, and power
consumption. The primary findings are summarized below.

The analysis of the on-off modulation showed that the tether tip draws
a closed loop on a surface that is a combination of two planes and a
saddle surface. The tether tip moves along planar orbits when the
tether voltage is off (free motion) and transits from one plane to the
other on the saddle surface when the voltage is on (forced motion,
thrusting arc). The motion was solved as a function of given length of
the trusting arc, sail angle with respect to the Sun direction, and
ratio of the electric solar wind sail thrust to the centrifugal force.

The transverse thrust efficiency of the smooth modulation combines two
effects. One is the electric sail force geometry and the other is the
tether voltage modulation. The thrust is proportional to the solar wind
component perpendicular to the tether. Thus the instantaneous thrust
direction changes from being fully radial (tether normal to the
orbital plane) to having both radial and transverse components defined
by the sail angle (tether in orbital plane). The crest-to-trough ratio
of the voltage modulation increases with increasing tether coning
angle. The thrust averaged over the rotation phase leads then to an
efficiency that decreases strongly depending on the coning angle. For
a coning angle close to 0$^{\circ}$ (7$^{\circ}$), the transverse
thrust is about 0.23 (0.17) times the total thrust at the sail angle of
45$^{\circ}$. This means that a fast spinning sail is more efficient
implying the importance of the tether material tensile strength for
the sail efficiency.

The on-off modulation was motivated by the fact that the transverse
thrust is mostly produced in relatively narrow sectors around the
orbital plane. Having the tether voltage fully on only along these
thrusting arcs, most of the transverse thrust is captured. Having the
voltage off elsewhere reduces the radial thrust. As the result, the
transverse thrust generated is about 0.12 times the total thrust
available in a wide range of the sail angle and electric sail force to
centrifugal force ratio ($\rho$). Thus, for realistic $\rho$ values,
it approaches the smooth mode efficiency values. Furthermore, the
power consumption is 0.13 times the smooth modulation
consumption. Thus the transverse thrust efficiency scaled to the power
system mass (proportional to power consumption) is about 0.8 for the
on-off mode and 0.23 for the smooth mode. Finally, the thrust angle of
the on-off mode approximately equals the sail angle. For the smooth
mode, the thrust angle is about half of the sail angle with the
maximum of about 20$^{\circ}$ reached at 45$^{\circ}$ after which the
thrust angle is constant for sail angles less than 60$^{\circ}$. The
numbers given here are for a thrusting arc length of 45$^{\circ}$.


\appendix

\section{}
\label{app:a}

The equation of motion Eqs. (\ref{eq:lambdamotion})--(\ref{eq:phimotion})
can be solved for the on-off modulation as follows. The orbital change
associated with the flip of the free plane takes place near the plane
of $Z = 0$ for arcs A and B (Fig. \ref{fig:taco}) where the coning
angle $\Lambda \approx 0$. Thus the equation of motion
Eqs. (\ref{eq:lambdamotion})--(\ref{eq:phimotion}) can be linearized
in $\Lambda$ around both $\phi \approx 0$ and $\phi \approx \pi$. The
equation of motion can then be written as
\begin{eqnarray}
\ddot{\Lambda} + \Lambda\dot{{\phi}}^2 &=& - k\left(p\Lambda\sin\alpha\cos\phi+\cos\alpha\right) 
\label{eq:linear_a}\\
\ddot{\phi} &=& - pk\sin\alpha\sin\phi,
\label{eq:linear_b}
\end{eqnarray}
where we have added factor $p=\pm 1$ to include both arcs A ($p=1$)
and B ($p=-1$) in the analysis while further expansions in $\phi$ can
be considered at $\phi = 0$ for both arcs. Similarly, the plane of
free motion of Eq. (\ref{eq:freeplane}) is linearized to read as
\begin{equation}
\Lambda + \tan\mu\sin\phi = 0
\label{eq:linearfreeplane}
\end{equation}
for both arcs since for the arc B, the flipping of the plane of the
free rotation ($\mu \rightarrow -\mu$) and expansion around $\phi
\approx \pi$ ($\sin(\phi+\pi) = -\sin\phi$)
cancels. 

Next, Eq. (\ref{eq:linear_b}) can be solved to read as
\begin{equation}
\dot\phi^2 = \nu_{\Theta}^2 + 2pk\sin\alpha(\cos\phi - 1), 
\label{eq:dotvarphi}
\end{equation}
where the constant of integration ($\nu_{\Theta}^2$) is such that the
angular frequency equals to $\nu_\Theta$ at $\phi = 0$. The subscript
$\Theta$ denotes the tether thrusting arc A or B. Using the chain rule
($\ddot\Lambda = (d^2\Lambda/d\phi)\dot\phi^2 + (d\Lambda/d\phi)\ddot\phi$) in
Eq. (\ref{eq:linear_a}) and replacing both $\dot\phi^2$ and $\ddot\phi$
with Eq. (\ref{eq:dotvarphi}) and Eq. (\ref{eq:linear_b}),
respectively, differential equation for $\Lambda(\phi)$ can be written
as
\begin{eqnarray}
&&\frac{d^2\Lambda}{d\phi^2}[1+2p\kappa_{\Theta}\sin\alpha(\cos\phi - 1)] - \frac{d\Lambda}{d\phi}p\kappa_{\Theta}\sin\alpha\sin\phi\nonumber\\
&+&\,\Lambda[1+p\kappa_{\Theta}\sin\alpha(3\cos\phi-2)+p\kappa_{\Theta}\sin\alpha] + \kappa_{\Theta}\cos\alpha = 0,
\label{eq:diffL}
\end{eqnarray}
where $\kappa_{\Theta} = k/\nu_{\Theta}^2$. 

As a next step, we replace $\sin\phi$ and $\cos\phi$ by their power
expansions in Eqs. (\ref{eq:dotvarphi}) and (\ref{eq:diffL}). First,
$\dot\phi^2$ can be readily written as
\begin{equation}
\dot\phi^2 \approx \nu_{\Theta}^2 - pk\sin\alpha\phi^2 + \mathcal{O}{(\phi^4)}.
\label{eq:Fserie}
\end{equation}
Then $\Lambda(\phi)$ can be solved as a power series as
\begin{equation}
\Lambda = \sum_{i=0}^N a_i\phi^i.
\end{equation}
The coefficients $a_i$ can be obtained by a computer algebra systems
such as Maxima \cite{maxima}, and $\Lambda(\phi)$ can be written as
\begin{equation}
\Lambda \approx \Lambda_{\Theta} + a^{\Theta}_2\phi^2 + \mathcal{O}(\phi^4),
\label{eq:Lserie}
\end{equation}
where
\begin{equation}
a^{\Theta}_2 = - \frac{1}{2}\left(\Lambda_{\Theta}(1 + 2p\kappa_{\Theta}\sin\alpha) + \kappa_{\Theta}\cos\alpha\right)
\label{eq:a2}
\end{equation}
and is given in terms of the free coefficient $a^{\Theta}_0 =
\Lambda_{\Theta}$ to be determined by the continuity conditions at
$\phi = \pm\phi_{\Theta}$. As it can be expected, $\Lambda(\phi)$ is symmetric as
\begin{equation}
a^{\Theta}_3 = - \frac{1}{6}a^{\Theta}_1(1+p\kappa_{\Theta}\sin\alpha) = 0
\end{equation}
since $a^{\Theta}_1 = \dot\Lambda_{\Theta} = 0$ when $\phi = 0$.  For
the free motion, $\lambda(\phi)$ in Eq. (\ref{eq:linearfreeplane}) and
$\dot\phi$) in Eq. (\ref{eq:freedotphi}) read as
\begin{equation}
\Lambda \approx -\tan\mu(\phi - \textsc{\small 1/}_{\textsc{\small 6}}\,\phi^3+ \mathcal{O}{(\phi^5)})
\end{equation}
and
\begin{equation}
\dot\phi^2 \approx \frac{\omega_0^2}{1+\tan^2\mu}(1+2\tan^2\mu\phi^2+\mathcal{O}{(\phi^4)}) \approx \omega_0^2
\label{eq:dotphi_app}
\end{equation}
as expanded in $\phi$. In Eq. (\ref{eq:dotphi_app}), $\dot\phi^2
\approx \omega_0^2$ when only the linear terms in $\tan\mu$ are
considered.

To solve $\Lambda_{\Theta}, \nu_{\Theta}, \tan\mu$, and $\phi_B$ in
terms of the free parameter $\phi_A$ the continuity equations for
$\Lambda$, $\dot\Lambda$ and $\dot\phi^2$ are written. As
$\Lambda(\phi)$ and $\dot\phi(\phi)$ are symmetric at origin, these equations
read as
\begin{eqnarray}
&&\Lambda_{\Theta} + a^{\Theta}_2\phi_{\Theta}^2 = - \tan\mu(\phi_{\Theta} - \textsc{\small 1/}_{\textsc{\small 6}}\,\phi_{\Theta}^3)\label{eq:C_Lambda}\\
&&2a^{\Theta}_2\phi_{\Theta} = - \tan\mu(1 - \textsc{\small 1/}_{\textsc{\small 2}}\,\phi_{\Theta}^2)\label{eq:C_Lambdadot}\\
&&\nu_{\Theta}^2 - pk\sin\alpha\phi^2_{\Theta}  = \omega_0^2.
\label{eq:C_nudot}
\end{eqnarray}
For both arcs A and B, $\nu_{\Theta}^2$ and $\Lambda_{\Theta}$ can be solved in terms of $\tan\mu$ and $\phi_{\Theta}$ as
\begin{equation}
\nu_{\Theta}^2 = \omega_0^2 + p k \sin\alpha\phi_{\Theta}^2
\label{eq:nua}
\end{equation}
and
\begin{equation}
\Lambda_{\Theta} = -\frac{1}{2}\tan\mu(1+\textsc{\small 1/}_{\textsc{\small 6}}\,\phi_{\Theta}^2)\phi_{\Theta}.
\label{eq:LambdaT}
\end{equation}

For the arc A ($p = 1$), the arc half length ($\phi_A$) and the
angular speed ($\omega_0$) are free parameters, $\nu_A$ and $\tan\mu$,
and $\Lambda_A$ are solved. After solving $\tan\mu$, $\nu_A$ and $\Lambda_A$
are available from Eq. (\ref{eq:nua}) and Eq. (\ref{eq:LambdaT}).
\begin{eqnarray}
\tan\mu &=& \frac{12\kappa_0\cos\alpha \phi_A}{12 + 24\kappa_0\sin\alpha\phi_A^2 + (1 + 2\kappa_0\sin\alpha)\phi_A^4 + \kappa_0\sin\alpha\phi_A^6}\label{eq:tanmu}\\
&\approx& \kappa_0\cos\alpha\phi_A(1-2\kappa_0\sin\alpha\phi_A^2 + \mathcal{O}{(\phi^4)}),
\end{eqnarray}
where $\kappa_0 = k/\omega_0^2$ and Eq. (\ref{eq:tanmu}) gives the
exact continuity at $\phi = \phi_A$.

For the arc B ($p = -1$), $\nu_0$ and $\mu$ are known, and $\phi_B$,
$\Lambda_B$, and $\nu_B$ are solved. This case is more complicated
since Eqs. (\ref{eq:C_Lambda}), (\ref{eq:C_Lambdadot}), and
(\ref{eq:C_nudot}) are non-linear in $\phi_B$. To solve these
equations, we assume that $\phi_A = \phi_B + \delta\phi$, expand the
equations in $\delta\phi$ to solve $\delta\phi$ to read as
\begin{eqnarray}
\delta\phi &=& - \frac{\kappa_0\sin\alpha(48\phi_A^3 + 4\phi_A^5 + 2\phi_A^7)}{12 + 72\kappa_0\sin\alpha\phi_A^2 + (10\kappa_0\sin\alpha - 3)\phi_A^4 + 7\kappa_0\sin\alpha\phi_A^6}\\
 &\approx& -4 \kappa_0\sin\alpha\phi_A^3(1 + \textsc{\small 1/}_{\textsc{\small 12}}\,(1 - 24 \kappa_0\sin\alpha)\phi_A^2 + \mathcal{O}{(\phi^4)}).
\label{eq:deltaphi}
\end{eqnarray}
Finally, the half arc length $\phi_B$ can be written as
\begin{equation}
\phi_B = \phi_A(1-4\kappa_0\sin\alpha \phi_A^2 + \mathcal{O}{(\phi^4)}).
\label{eq:phiB}
\end{equation}

\section{}
\label{app:b}

The definite integrals used in this study are given below. They can be
deduced by partial integration, integral tables, or computer algebra
systems.

\begin{equation}
\frac{1}{2\pi}\int_0^{2\pi}\frac{d\phi}{(1+\chi\cos\phi)^2} = \frac{1}{(1-\chi^2)^{\frac{3}{2}}}
\label{eq:integrala}
\end{equation}

\begin{equation}
I_0 = \frac{1}{2\pi}\int_0^{2\pi}\frac{d\phi}{(1+\chi\cos\phi)^3} = \frac{(2+\chi^2)}{2(1-\chi^2)^{\frac{5}{2}}}
\end{equation}

\begin{equation}
I_1 = \frac{1}{2\pi}\int_0^{2\pi}\frac{\cos\phi d\phi}{(1+\chi\cos\phi)^3} = -\frac{3\chi}{2(1-\chi^2)^{\frac{5}{2}}}
\end{equation}

\begin{equation}
I_2 = \frac{1}{2\pi}\int_0^{2\pi}\frac{\cos^2\phi d\phi}{(1+\chi\cos\phi)^3} = \frac{(1+2\chi^2)}{2(1-\chi^2)^{\frac{5}{2}}}
\end{equation}



\section*{Acknowledgements}
This work was supported by Academy of Finland.

\bibliographystyle{model6-num-names}
\bibliography{<your-bib-database>}



\end{document}